\documentclass[twocolumn,showpacs,preprintnumbers,amssymb]{revtex4}
\usepackage{epsfig,subfigure}
\usepackage{graphicx}
\usepackage{amsmath}
\usepackage{color}

\newcommand{\nuc}[3]{\ensuremath{ {}^{#2}_{#3}\mathrm{#1} }}

\newcommand{\LS}{L\mbox{-}S}

\begin{document}
\preprint{FIS-UI-TH-05-01}
\def\Journal#1#2#3#4{{#1} \textbf{#2}, #3 (#4)}
\def\RPP{{Rep. Prog. Phys.}}
\def\PRC{{Phys. Rev. C}}
\def\FP{{Found. of Physics}}
\def\ZPA{{Z. Phys. A}}
\def\NPA{{Nucl. Phys. A}}
\def\JPG{{J. Phys. G: Nucl. Part}}
\def\PRL{{Phys. Rev. Lett.}}
\def\PRpt{{Phys. Report.}}
\def\PLB{{Phys. Lett. B}}
\def\AP{{Ann. Phys. (N.Y.)}}
\def\EPJA{{Eur. Phys. J. A}}
\def\NP{{Nucl. Phys.}}
\def\ZP{{Z. Phys}}
\def\RMP{{Rev. Mod. Phys.}}
\input epsf

\title{Nilsson parameters $\kappa$ and $\mu$ in the relativistic mean field
models}

\author{A. Sulaksono, T. Mart}
\affiliation{Departemen Fisika, FMIPA, Universitas Indonesia, Depok
16424, Indonesia}

\author{C. Bahri}
\affiliation{Departemen Fisika, FMIPA, Universitas Indonesia, Depok
16424, Indonesia}
\affiliation{Department of Physics and Astronomy, Louisiana State
University, Baton Rouge, LA 70803, USA}

\begin{abstract}
Nilsson parameters $\kappa$ and $\mu$ have been studied in the framework of the
relativistic mean field (RMF) models. They are used to investigate the reason that  the RMF models give a relatively well prediction of the spin-orbit splitting, but fail to reproduce the placement of the
states with different orbital angular momenta. Instead of the relatively small effective mass $M^*$, the independence of $M^*$ from  the angular momentum $\emph{l}$ is found to be the reason.
\end{abstract}


\maketitle


\section{Introduction} 
The finite-range (FR) (see Refs.~\cite{wal,pg,serot,ring2} for a review) and
point-coupling (PC) (see Refs.~\cite{niko,rusnak,buerven} for a review) types of
relativistic mean-field (RMF) models have been successful to
describe the bulk properties as well as the deformation in a wide mass spectrum
of nuclei. The inter-connection of the models and their relations to
non-relativistic models, like the Skyrme Hartree-Fock (SHF) ones, have been established~\cite{anto1}.
The role of the exchange in RMF-FR~\cite{bosen,boersma,zhang,schmid,quelle} and RMF-PC~\cite{anto2} models for finite nuclei have been also explored.  

A significant attention has been paid to explore the role of the spin-orbit
potential ($V_{\LS}$) in RMF models for various problems and
applications that are connected to single particle spectra
predictions (see for examples
Refs.~\cite{pg,quelle,bender,jamin,koepf,baran,rusnak2,yosida,todd}). The standard parameter set of the Nilsson model has been also quite successful in reproducing single particle spectra of stable nuclei~\cite{tord}. New parameter sets were proposed to improve the  predictability of the model for neutron rich~\cite{zhang2} and proton rich~\cite{sun} nuclei. Single particle levels of this parameter sets are compared in Refs.~\cite{zhang2,sun} with those obtained by SHF and RMF models. We  also note that the origin of the prolate dominance shapes over the oblate ones can be explained in the framework of a Nilsson model as an effect of the strong interference between spin-orbit and orbit-orbit terms of the Nilsson potential~\cite{tajima}.  
So far, however, except for the pseudo-spin symmetry study in finite
nuclei~\cite{cb}, there have been no other investigations of the role
 of the orbit-orbit potential ($V_{LL}$) which is  directly derived from RMF models.

Odd nuclei and single particle spectra in an RMF-FR model using different level of approximations (spherical and deformed) were computed and compared in Ref.~\cite{rutz}. In Ref~\cite{quelle}, single particle splitting energies between spin-orbit partners along some isotonic chain (O, Ca, Sn) are also examined in the framework of RMF, SHF and relativistic Hartree-Fock models. Furthermore, there is another method to study spin-orbit potential by exploring high-spin data~\cite{satula}. With this method one can avoid the scarce and uncertain  data available on spin-orbit splittings and their isotopic as well as isotonic dependences. Here we quote from Ref.~\cite{satula}, for example, that the most recent experimental data evaluations \cite{oros} give $\Delta~\epsilon_{d_{3/2}-d_{5/2}} \approx$ 6 MeV  for  $\nuc{Ca}{40}{}$ and $\Delta~\epsilon_{d_{3/2}-d_{5/2}} \approx$ 5 MeV  for  $\nuc{Ca}{48}{}$,  while older works give $\Delta~\epsilon_{d_{3/2}-d_{5/2}} \approx$ 6.8 MeV \cite{swift}, 7.3 MeV \cite{tyren},  7.7 MeV\cite{ray} for $\nuc{Ca}{40}{}$ and  $\Delta~\epsilon_{d_{3/2}-d_{5/2}} \approx$ 5.3 MeV \cite{ray} for $\nuc{Ca}{48}{}$. More detailed information on single particle levels can be found in Ref.~\cite{Isakov}. Since the method is based on a direct comparison of the excitation energies of terminating states, the corelations beyond mean-field can be strongly suppresed. Nevertheless, this method is still constrained by the limited knowledge on the time-odd component of the nonrelativistic mean-field~\cite{satula}.

In this paper, we will revisit and study the single particle spectra (SPS) of
$\nuc{Pb}{208}{}$, $\nuc{Sn}{132}{}$ and $\nuc{Ca}{40}{}$ in the RMF models in order to understand the origin of their predictive powers for spin-orbit splitting and the reason that the relative placements
of the states with different orbital angular momenta $\emph{l}$ are not well
reproduced~\cite{rutz}.  Afterward, we will try to find the connection between their
SPS predictions with their effective masses ($M^*$) through their $V_{\LS}$ and
$V_{LL}$ potentials in $\nuc{Pb}{208}{}$. Spherically symmetric calculations are used due to the robustness of the
spectral differences against polarization effects~\cite{rutz}.

We choose NL-Z, NL-Z2, NL-VT1 (RMF-FR)~\cite{pg,bender,rufa} and PC-F1
(RMF-PC)~\cite{buerven} because they nearly have the same procedure
to adjust their coupling constants, hence the prediction bias due to
the different fitting procedure can be minimized.
\begin{figure*}
\centering
 \mbox{\epsfig{file=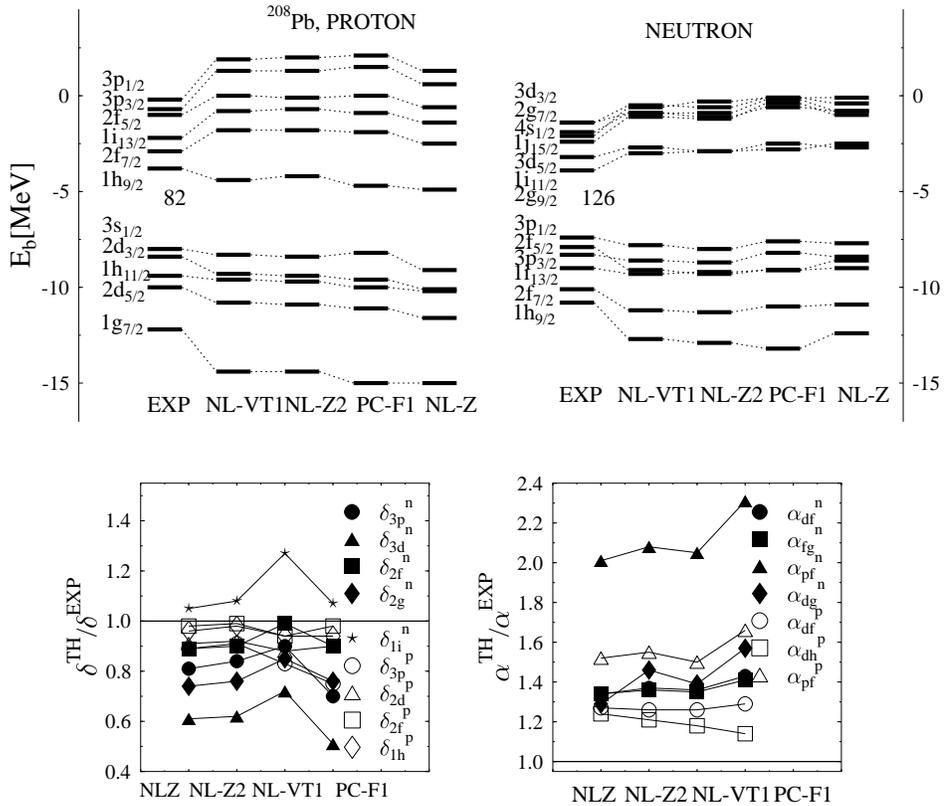,height=11.0cm}}
\caption{Single particle spectra (SPS) for $\nuc{Pb}{208}{}$  (top): left and right panels are for proton and neutron respectively. Spin-orbit splitting ($\delta$) and relative position between two different levels with different angular momenta ($\alpha$) of $\nuc{Pb}{208}{}$  are shown in the left and right bottom panels. Experimental data (EXP) are taken
  from Ref.~\cite{Isakov}. RMF models with NL-Z, NL-Z2, NL-VT1, and PC-F1 parametrizations are shown.\label{Pb208sps}}
\end{figure*}
\section{SPS predictions of RMF models}
In this section we will study the $\nuc{Pb}{208}{}$,  $\nuc{Sn}{132}{}$ and $\nuc{Ca}{40}{}$ SPS predictions of RMF
models. The experimental
single particle data are taken from~\cite{Isakov}.  The $\nuc{Pb}{208}{}$
neutron ($2f_{7/2}$ and $1h_{9/2}$) and proton ($2d_{5/2}$ and $1g_{7/2}$) data as well as the $\nuc{Ca}{40}{}$ neutron ($1d_{5/2}$) and proton  ($1d_{5/2}$) data
are obtained by averaging over the spectroscopic factors~\cite{Isakov}.
To analyze the relative position between two different levels with different  angular momenta ($\alpha$) we use the following
formulas~\cite{yosida}
\begin{eqnarray}
\bar{E}_l&=&\sum_{j=l-1/2}^{j=l+1/2}\frac{(2j+1)E_j}{2(2l+1)},~~~~~~
\alpha_{lk}~=~ \bar{E}_l -\bar{E}_k,
\label{alpha}
\end{eqnarray}
where $\bar{E}_l$ is the average energy of the spin-orbit partner with angular momentum $\emph{l}$. The relative position between $\emph{k}$ and $\emph{l}$  levels can be determined from the difference between $\bar{E}_l$ and $\bar{E}_k$ ($\alpha_{lk}$).

Figure~\ref{Pb208sps} shows the SPS for $\nuc{Pb}{208}{}$. In the top panels, it can be seen that all models have a similar trend in placing and ordering the proton
and neutron single particle energies.  The gap between an occupied and an unoccupied levels is relatively well reproduced for proton but  quite poor in the case of neutron. Compared with experimental
data, the SPS lines do not coincide.  The deviations show up
significantly in the $1g_{7/2}$, $3p_{3/2}$ and $3p_{1/2}$ proton states and  in the $1h_{9/2}$ neutron state.  For the neutron spectrum, similar to
Ref.~\cite{rutz},  the ordering is reversed between $1i_{13/2}$,
$3p_{3/2}$ and $2f_{5/2}$ states. There are quite
significant discrepancies in the spacing between $1i_{11/2}$ and $3d_{5/2}$ states, as well as between  $1h_{9/2}$ and $2f_{7/2}$  states, with
experimental data. For the proton spectrum, it occurs between $1h_{9/2}$ and $2f_{7/2}$ states as well as between $2d_{5/2}$ and $2g_{7/2}$ states.

The trends of the spin-orbit splitting (see the lower-left panel of Fig.~\ref{Pb208sps}) and the
relative position of SPS (see the lower-right panel of Fig.~\ref{Pb208sps}) of NL-Z are similar to the case of NL-Z2 but
different from those of NL-VT1 and PC-F1. For proton, except for the
splitting in $3p$
states of PC-F1, all parameter sets have only 15\%
deviation from their experimental values.
Nevertheless, since the positions of $3p$ states are quite far from the Fermi surface,
we can say that all parameters sets give a good prediction of the
proton spin-orbit splitting.  In the case of neutron, only the splitting of $2f$ states deviates by 
less than 10\% from the experimental value.  NL-VT1 has four gaps with
deviations in splitting less than 20\%. Unfortunately, it has a gap ($1i$
states) with a  more than 20\% deviation and the gap is larger  than the corresponding
experimental data. The position of these states is 
above the Fermi surface. NL-Z2 and NL-Z have more or less 20\%
deviation in the splitting of $3p$ states (the position of these states is around Fermi surface)
and for PC-F1 the deviation of that splitting is larger than  20\%. A quite large
deviation appears in the splitting of $3d$ and $2g$ states (the positions of both spin-orbit partners are above Fermi
surface). It seems that all parameters sets are unable to
give good predictions in the neutron spin-orbit splitting.
The SPS relative position  of proton has a better
prediction than that of neutron.  Proton has two $\alpha$s which have 
deviation less than  30\% and one $\alpha$ above the Fermi level ($\alpha_\mathit{pf}$).
 We also note that $\alpha_\mathit{pf}$ has 60\% deviation. Neutron has three $\alpha$s
with deviations from experimental data between  30\% and 50\% and it
has even one $\alpha$ ($\alpha_{pf}$) that deviates by about 100\% from
experimental data.  The positions of those states ($3p$ and $2f$) are around the Fermi surface. 
\begin{figure*}
\centering
 \mbox{\epsfig{file=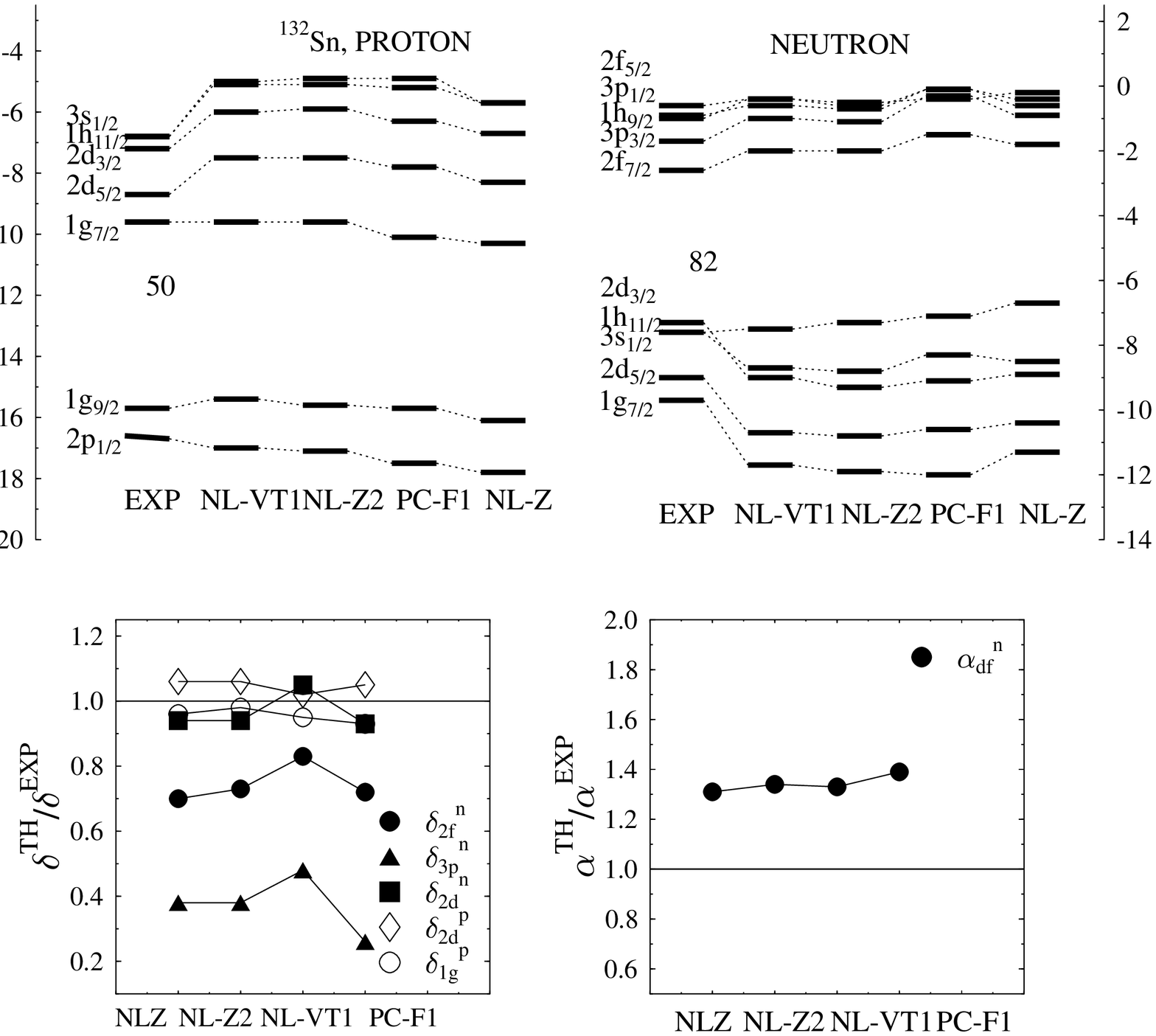,height=11.0cm}}
\caption {Same as Fig.~\ref{Pb208sps}, but for $\nuc{Sn}{132}{}$. \label{Sn132sps}}
\end{figure*}

Figure~\ref{Sn132sps}  shows the SPS for $\nuc{Sn}{132}{}$. In the top panels, it can be seen  that all models have similar trend in levels placing and ordering.  Their SPS lines do not coincide  with experimental data. For the neutron spectrum, the reversed ordering between $2d_{3/2}$ with $3s_{1/2}$ and $2h_{11/2}$ states occurs. A significant difference in spacing with experimental data occurs between  $2d_{5/2}$ and $1g_{7/2}$ states of proton. The bottom panels show that all parameter sets predict acceptable spin-orbit splittings for proton but not for neutron. The relative position between $2d$ and $2f$ levels for neutron ($\alpha_{df}^n$) deviates by almost 40 \% from experimental data. The trends of the spin-orbit splitting (see the lower-left panel of Fig.~\ref{Sn132sps}) and the
relative position of the SPS (see the lower-right panel of Fig.~\ref{Sn132sps}) of NL-Z are similar to those of NL-Z2 but
different from those of NL-VT1 and PC-F1. 
\begin{figure*}
\centering
 \mbox{\epsfig{file=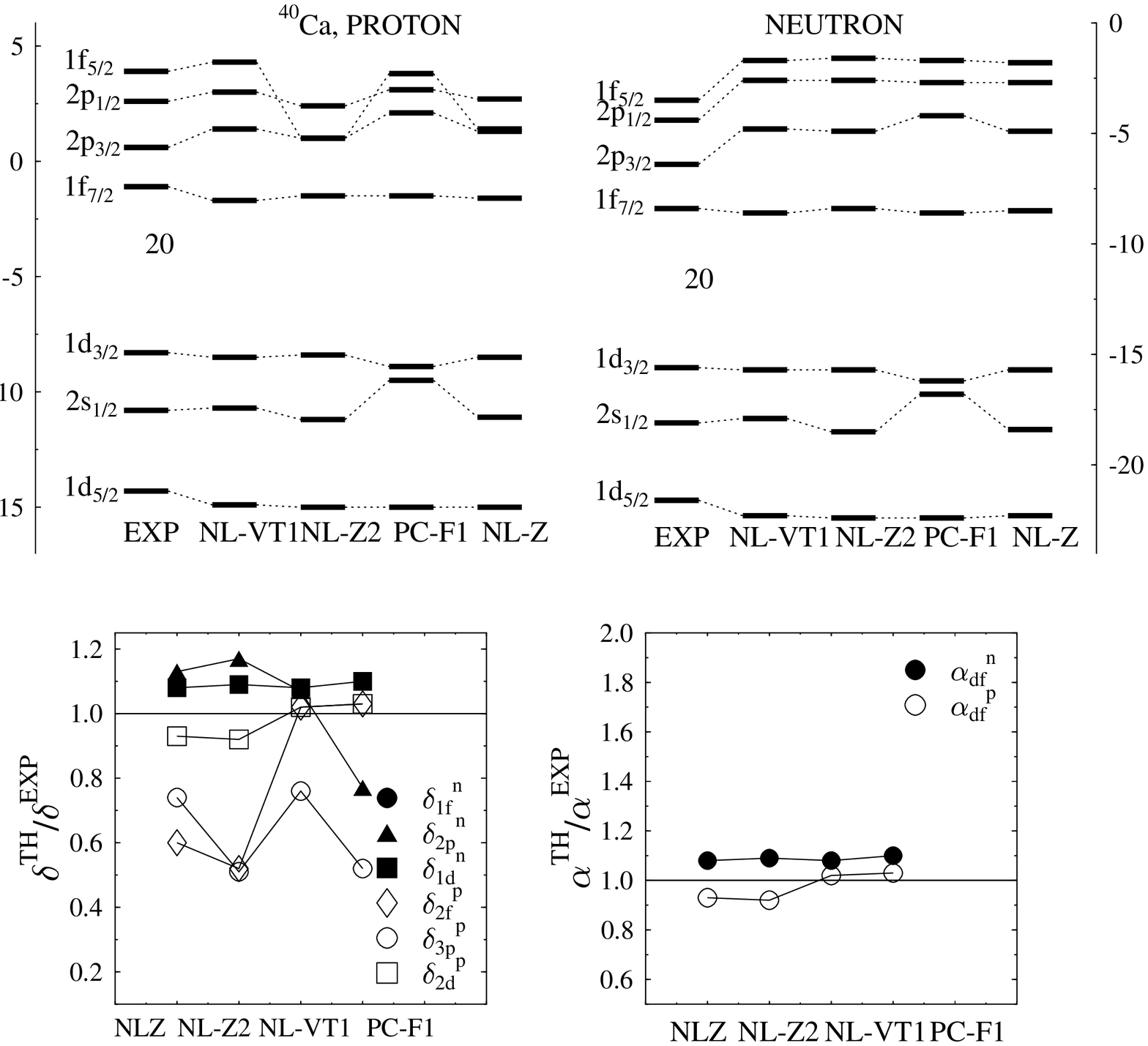,height=11.0cm}}
\caption{Same as Fig.~\ref{Pb208sps}, but for $\nuc{Ca}{40}{}$. \label{Ca40sps}}
\end{figure*}

Figure~\ref{Ca40sps} shows the SPS for $\nuc{Ca}{40}{}$. In the upper-left panel (proton) it is shown that NL-VT1 reproduces experimental values of every single particle energy. PC-F1 has a too narrow spacing between $1d_{3/2}$ and $2s_{1/2}$ states. The reversed ordering of $2f_{5/2}$ and $2p_{1/2}$ states occurs in the case of NL-Z and NL-Z2. In the upper-right panel (neutron) it can be seen that each model does not really have a similar trend in levels placing and ordering. Similar to the case of proton,  here PC-F1 has a too narrow spacing between $1d_{3/2}$ and $2s_{1/2}$ states. The lower panels show that, in contrast to  $\nuc{Pb}{208}{}$ and  $\nuc{Sn}{132}{}$,  $\nuc{Ca}{40}{}$ has a better spin-orbit splitting for neutron rather than for proton. The relative positions between $1d$ and $1f$ levels for neutron and proton ($\alpha_{df}^n$) deviate by less than 20 \% from experimental data. 

These results confirm the findings of Ref.~\cite{rutz} that the relative placement and ordering of the
states in RMF models are not well reproduced.
In addition, we note that RMF-FR (NL-Z, NL-Z2 and NL-VT1)  has a better prediction than RMF-PC (PC-F1)  and
the presence of the tensor terms (NL-VT1) enhances the improvements in 
placing and ordering of single particle states, particularly in lighter nuclei (e.g., $\nuc{Ca}{40}{}$). However, these improvements  are still not adequate to overcome the problem. 
\\

\section{Nilsson Parameters $\kappa$ and $\mu$ of RMF Models}
It has been known that all models presented here have
$M^*/m \approx 0.6$.  Unlike the non-self-consistent calculations (models using
Wood-Saxon or Nilsson potentials), where the SPS has a direct connection with the
potential parameters, the connection is not so obvious in the RMF models
because it is hidden by the self consistency condition.  Therefore, it is natural to translate $M^*/m$ of RMF models into $V_{\rm c}$, $V_{LL}$ and $V_{\LS}$
by taking a non-relativistic limit, where they resemble a Wood-Saxon or Nilsson potential.  The interpretation of the results of this section
will be given in the next section by varying $M^*/m$ in one model and
studying its SPS prediction for $\nuc{Pb}{208}{}$.

\begin{figure}
\centering
 \mbox{\epsfig{file=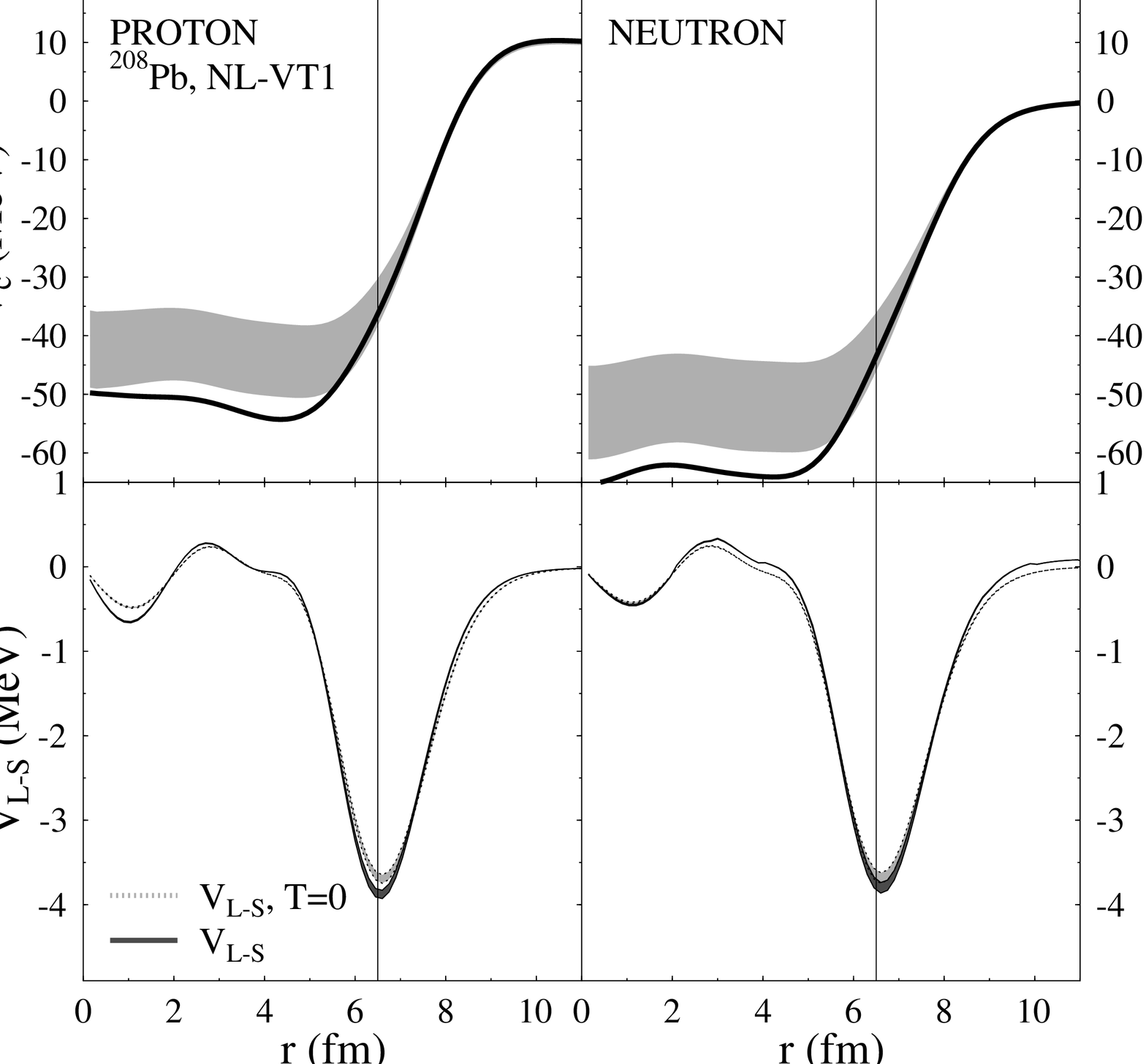,height=7.4 cm}}
\caption{Shaded areas are the energy-dependent central potential $V_{\rm c}$ (top) and the spin-orbit  potential $V_{L-S}$ (bottom). The proton and
  neutron parts are shown in the left and right panels, respectively. $R=1.1\,A^{1/3}$ fm is the Fermi surface radius. Solid lines in the top  are obtained from the summation of the scalar and time components of vector potentials (the dominant term in $V_{\rm c}$). NL-VT1 parameterization is used. $T = 0$ (bottom) means that the tensor potential is turned off.
  \label{potcomp}}
\end{figure}

The Hamiltonian of the RMF model in spherical systems is
\begin{eqnarray}
H = \vec{\alpha}\cdot (\vec{p}+i\gamma_0 \vec{T})+\gamma_0 (m+S) +V_0,
\label{eq:hmtn}
\end{eqnarray}
where $H \Psi_k^{\pm}$= $\epsilon_k^{\pm}\Psi_k^{\pm}$ is fulfilled.
Using the general convention for $\Psi_k^+$, i.e.,  $\Psi_k^+$ =
$\begin{pmatrix} g_k \chi_{k}^{m_j}\\ i f_k \chi_{-k}^{m_j}
\end{pmatrix}$, the positive energy equation for
the upper component becomes
\begin{widetext}
\begin{eqnarray}
\biggl[\partial_r^2&+&\frac{2}{r}\partial_r -\frac{\vec{L}^2}{r^2}-\left(
\frac{(\partial_r \Delta)}{(2 m +\epsilon_k^{+'}-\Delta)} - 2 T_r\right)
\frac{\vec{\sigma}\cdot \vec{L}}{r}+\frac{(\partial_r
  \Delta)}{(2 m +\epsilon_k^{+'}-\Delta)}\partial_r\nonumber\\&+&\left(\frac{2
  T_r}{r}-T_r^2+(\partial_r T_r)+\frac{(T_r\partial_r \Delta)}{(2 m +\epsilon_k^
  {+'}-\Delta)}\right)+(\epsilon_k^{+'}-\Sigma)(2m+\epsilon_k^{+'}-\Delta
  )\biggr] g_k=0,\label{eq:uppere+}
\end{eqnarray}
\end{widetext}
with $\Sigma= S + V_0$, $\Delta = V_0 - S$, and $\epsilon_k^{+'}=
\epsilon_k^{+}- m$, while $S$, $V_0$ and $\vec{T}$ indicate the scalar, time-component of the vector and tensor potentials, respectively. 
The non-relativistic form of Eq.~(\ref{eq:uppere+}) can be derived. The
Darwin term $\left(\frac{(\partial_r
  \Delta)}{(2 m +\epsilon_k^{+'}-\Delta)}\partial_r\right)$ in Eq.~(\ref{eq:uppere+})  can be absorbed by transforming the $g_k$ wave function into
the new one, $G^{+}_k$ ~\cite{jamin}. This leads to a Schr\"odinger form, i.e.,
\begin{eqnarray}
\biggl(\frac{p_r^2}{2m}+\frac{\boldsymbol{L}^2}{2m r^2}+V_{\rm c}(r,\epsilon^{+'})&+&
V_{\LS}(r,\epsilon^{+'})\frac{\boldsymbol{\sigma}\cdot\boldsymbol{L}}{r}\biggr)G^{+}_k
\nonumber\\&=&\epsilon^{NR} G^{+}_k,\label{eq:nrfedep}
\end{eqnarray}
where
$\epsilon^{NR}=\epsilon^{+'}\left(1+\frac{\epsilon^{+'}}{2m}\right)$~\cite{jamin}. In
finite nuclei, the second term in $\epsilon^{NR}$ is smaller than one,
and
\begin{eqnarray}
 V_{\LS}(r,\epsilon^{+'})&=&\frac{1}{2 m}\left[\frac{(\partial_r\Delta)}{
    (2m +\epsilon^{+'}-\Delta) }- 2T_r\right],
\end{eqnarray}
\begin{widetext}
\begin{eqnarray}
  V_{\rm c}(r,\epsilon^{+'})&=&\Sigma -\frac{\Sigma \Delta}{2m}+\frac{(\Sigma+\Delta)
  \epsilon^{+'}}{2 m}
    +\frac{3}{8}\frac{{(\partial_r \Delta)}^2}{m {(2m
      +\epsilon^{+'}-\Delta)}^2}
  +\frac{1}{2}\frac{(\partial_r \Delta)}{m
    (2m +\epsilon^{+'}-\Delta) r}\nonumber\\&&+\frac{1}{4}\frac{(\partial_r^2
    \Delta)}{m (2m +\epsilon^{+'}-\Delta)}
  +\frac{1}{2 m}\left[\frac{-2 T_r}{r}+T_r^2- \partial_r T_r -
    \frac{(T_r ~ \partial_r \Delta)} {(2m +\epsilon^{+'}-\Delta)}\right].
\end{eqnarray}
\end{widetext}
In heavy nuclei, the mean-field central potential $V_{\rm c}$ and the spin-orbit  potential $V_{\LS}$ are closer
to the non-relativistic results obtained by using the Wood-Saxon
potential~\cite{blokhin}.  Unlike the Wood-Saxon results, the RMF model has
a strong energy-dependent $V_{\rm c}$ and a weak energy-dependent $V_{\LS}$
(Fig.~\ref{potcomp}).  In $\nuc{Pb}{208}{}$, $V_{\rm c}^n$ is deeper than $V_{\rm c}^p$ due
to the fact that more neutrons are present rather than protons. $V_{\LS}$ of RMF model (NL-VT1) is deeper than
the Wood-Saxon prediction~\cite{blokhin}.
The tensor term gives only a minor additional contribution in spin-orbit potential near the
Fermi surface. There is no significant difference between neutron and
proton $V_{\LS}$ of $\nuc{Pb}{208}{}$ in the RMF model (NL-VT1). 

The study of non-relativistic potentials of RMF models has been done in many places
with different intentions and different methods to obtain $V_{\rm c}$ and
$V_{\LS}$~\cite{jamin,koepf,bender,pg,baran,rusnak2,yosida,quelle}.
However, a non-relativistic form of the RMF model like the Nilsson
one has not yet been explored, especially in analyzing the SPS. The
advantage of using the Nilsson model is that $V_{LL}$ could be
employed to analyze the relation between states of different $l$ and
their orderings.  Reference~\cite{cb} calculates $V_{LS}$ and $V_{LL}$
from several RMF models by using some different approximations  to
study the origin of the pseudo-spin symmetry. We note that the maximum value of $\epsilon^{+'}$ has the same order of magnitude with the difference value  between the scalar and the time components of vector potentials ($\Sigma$), e.g., in $\nuc{Pb}{208}{}$ neutron, the corresponding value is around 50 MeV. On the other hand,  $\Delta$ is a summation of the scalar and the time components of the vector potential and the corresponding maximum value for $\nuc{Pb}{208}{}$ neutron is more or less 400 MeV. It is also known that the nucleon mass is around 1000 MeV~\cite{pg}. Therefore to obtain a ``Nilsson
form''  we can use the assumption that
$\epsilon^{+'} < \Delta < m$, so that
$\epsilon^{NR}=\epsilon^{+'}\left(1+\frac{\epsilon^{+'}}{2m}\right) \approx
\epsilon^{+'}$ and ${(2m +\epsilon^{+'}-\Delta)}^{-1}$ $\approx$ ${(2m
  -\Delta)}^{-1}$.  After that, Eq.~(\ref{eq:nrfedep}) can be written as
\begin{widetext}
\begin{eqnarray}
 \left[\,\frac{p_r^2}{2m}+\frac{\boldsymbol{L}^2}{2m r^2}+V_{\rm c}^{\rm eff}(r)+V_{LL}^{\rm eff}(r)
  \boldsymbol{L}^2 + V_{\LS}^{\rm eff}(r)\boldsymbol{s}\cdot\boldsymbol{L}\,\right]\, G^{+}_k
  ~\approx~ \epsilon^{+'} G^{+}_k,\label{eq:nrfldep}
\end{eqnarray}
\end{widetext}
where $\boldsymbol{s}$=$\frac{1}{2} \boldsymbol{\sigma}$ and
\begin{eqnarray}
  V_{\rm c}^{\rm eff}(r)&=&\biggl[1+\frac{\Sigma+\Delta}{\{2m-(\Sigma+\Delta)\}}\biggr]
  \lim_{\epsilon^{+'}\rightarrow 0}  V_{\rm c}(r,\epsilon^{+'})\nonumber\\
  &&-\frac{1}{2m}\biggl[\frac{(\Sigma+\Delta)}{\{2m-(\Sigma+\Delta)\}}\biggr] ~ ~p_r^2,\nonumber
  \\
  V_{LL}^{\rm eff}(r)&=& \frac{1}{2m r^2}\biggl[
  \frac{(\Sigma+\Delta)}{\{2m-(\Sigma+\Delta)\}}\biggr],\nonumber\\
  V_{\LS}^{\rm eff}(r)&=& 2 \biggl[1+\frac{\Sigma+\Delta}{\{2m-(\Sigma+\Delta)\}}\biggr]
  \lim_{\epsilon^{+'}\rightarrow 0}\frac{V_{L-S}(r,\epsilon^{+'})}{r}.\nonumber\\
\end{eqnarray}
As another consequence, the energy-dependent $V_{\rm c}$ transforms into energy-independent $V_{\rm c}$
plus a small nonlocal term ($p_r$ dependent).  Equation~(\ref{eq:nrfldep}) can
be considered as the Nilsson form of the RMF model. Both
potentials can be compared with their partners from the Nilsson
model through Nilsson parameters $\kappa$ and $\mu$~\cite{ring}.

\begin{figure}
\centering
 \mbox{\epsfig{file=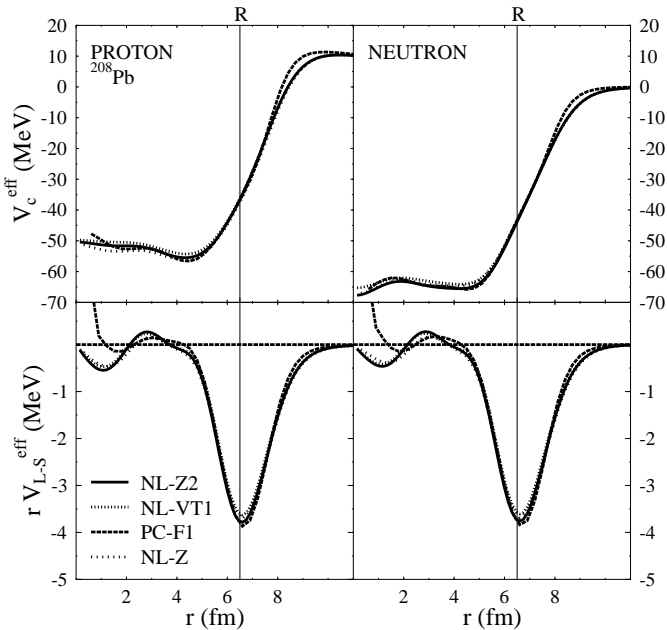,height=7.4 cm}}
\caption{The dominant part of the effective central potential $V_{\rm c}^{\rm eff}$ (top) and spin-orbit  potential  $r V_{L-S}^{\rm eff}$ (bottom). The proton and neutron parts are shown in the left and right panels, respectively. $R$=1.1$\,A^{1/3}$ fm is the Fermi surface radius. Here, RMF models with NL-Z, NL-Z2, NL-VT1, and PC-F1 parameterizations are employed. \label{poteff}}
\end{figure}

\begin{figure}
\centering
 \mbox{\epsfig{file=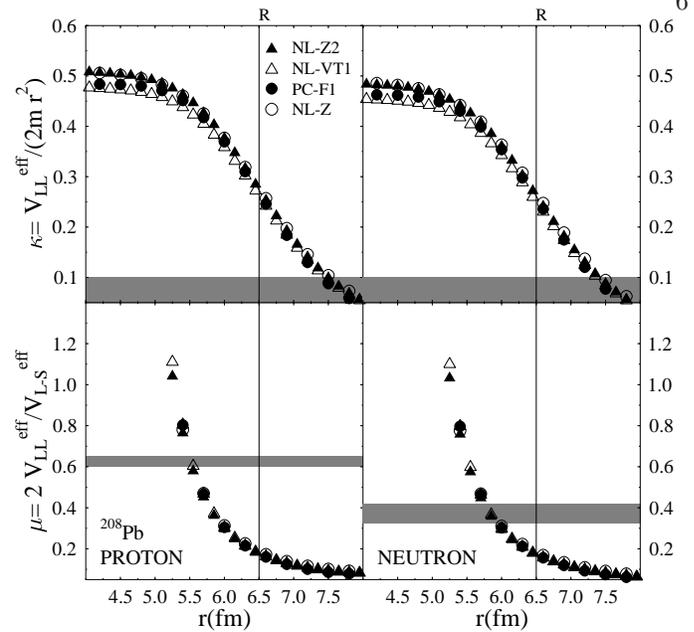,height=7.4 cm}}
\caption{Nilsson parameters $\kappa$ (top) and $\mu$ (bottom)  near the Fermi surface $R$. Left and right panels are for proton and neutron, respectively.\label{kamu}}

\end{figure}
The dominant parts of $V_{\rm c}^{\rm eff}$ and $V_{\LS}^{\rm eff}$ are taken merely to
compare the predictions among the presented models (parameter sets).
They have similar  $V_{\rm c}^{\rm eff}$ and $V_{\LS}^{\rm eff}$ predictions not only for
proton but also for neutron, especially near the Fermi surface.  Small
differences appear in the region around the center of nuclei and a small $V_{\rm c}^{\rm eff}$ deviation
also appears in the unoccupied region of PC-F1. Thus,  the
forms of $V_{\rm c}^{\rm eff}$ and $V_{\LS}^{\rm eff}$ are essentially almost model-independent
(see Fig.~\ref{poteff}).

In the Fermi surface, all parameter sets have similar  $\kappa$ and
$\mu$ for proton and neutron.  On the contrary, Nilsson model has different $\mu$ for proton
and neutron (see Fig.~\ref{kamu}).  Compared with the Nilsson model (shaded regions) RMF ones have larger
$\kappa$ but smaller $\mu$. The differences among
all parameter sets (models) in $\kappa$ only appear in the region
close to the center of nuclei.  It means that $\kappa$ and $\mu$ of RMF models can
be considered as model (parameter set) independent. Unlike in the Nilsson
model, where $\kappa$ and $\mu$ are independent from the position ($r$),
in RMF models both quantities depend on $r$. The spatial dependence of $V_{\LS}^{\rm eff}$ originates mainly from the energy-dependent potential $V_{\rm c}$. It should be noted that the non-relativistic model, like SHF, does not have such dependence. Therefore, we can consider this spatial dependence as a genuine feature of self-consistent RMF models.

\section{Interpretation}
\begin{table}[t]
\centering
\caption {Numerical values of coupling constants used in the parameter sets. Except for the NL-Z parameterization~\cite{rufa}, these values are adjusted with respect to $ M^*/m$. }\label{tab:params3}
\begin{ruledtabular}
\begin{tabular}{crrrrr}
Parameter &NL-Z & P-0.67& P-0.70 &P-0.75 &P-0.80 \\\hline
$g_S$ &10.06& 8.91& 8.45 &7.58 &7.22   \\
$g_V$ &12.91& 11.02& 10.26 &8.73 &7.94   \\
$g_R$ &9.69& 9.69& 9.69 &9.69 &9.69 \\\hline
$b_2$ &-13.51& -13.44& -13.41 &-13.06 &-13.49  \\
$b_3$ &-40.22& -29.74& -24.48 &-3.69 &30.07  \\\hline
$m_S$ &488.67& 488.67& 488.67 &488.67 &488.67  \\
$m_V$ &780& 780& 780 &780 &780  \\
$m_R$ &763& 763& 763 &763 &763 \\
\end{tabular}\\
\end{ruledtabular}
\end{table}

\begin{table}[b]
\centering
\caption {Nuclear matter properties predicted by NL-Z, P-067, P-0.70, P-0.75, and P-0.80 parameterizations. }\label{tab:NMPm*}
\begin{ruledtabular}
\begin{tabular}{lrrrrr}
 Parameter &NL-Z & P-0.67&P-0.70 &P-0.75 &P-0.80 \\\hline
$E/A$ (MeV)&-16.18& -16.30& -16.38 &-16.34 &-15.85   \\
$\rho_{nm}$ (fm$^{-3}$)&0.15& 0.16& 0.16 &0.17 &0.16 \\
$M^*/m$ &0.58& 0.67& 0.70 &0.75 &0.80  \\
$a_4$ ( MeV)&41.7& 41.1& 42.0 &42.8 &39.7  \\
\end{tabular}\\
\end{ruledtabular}
\end{table}

We prepared two variations of parameter sets to properly interpret the results. First, we varied the scalar coupling constant $g_s$ of NL-Z until
we obtained the desired $M^*/m$, while the four parameter sets were kept
constant.  Second, we fit the four parameter sets with $M^*/m$ varied.  The
parameter sets in the second procedure were obtained by fitting the four parameter
sets into the same observable that were used in obtaining the NL-Z parameter
set until they can reproduce nuclear matter properties.  As in Ref.~\cite{pg}
the $\chi^2$ becomes worse when $M^*/m$ is larger than 0.6.
The fitted parameter sets are
tabulated in Table ~\ref{tab:params3}, whereas their nuclear matter properties can
be seen in Table~\ref{tab:NMPm*}.

\begin{figure}[t]
\centering
\mbox{\epsfig{file=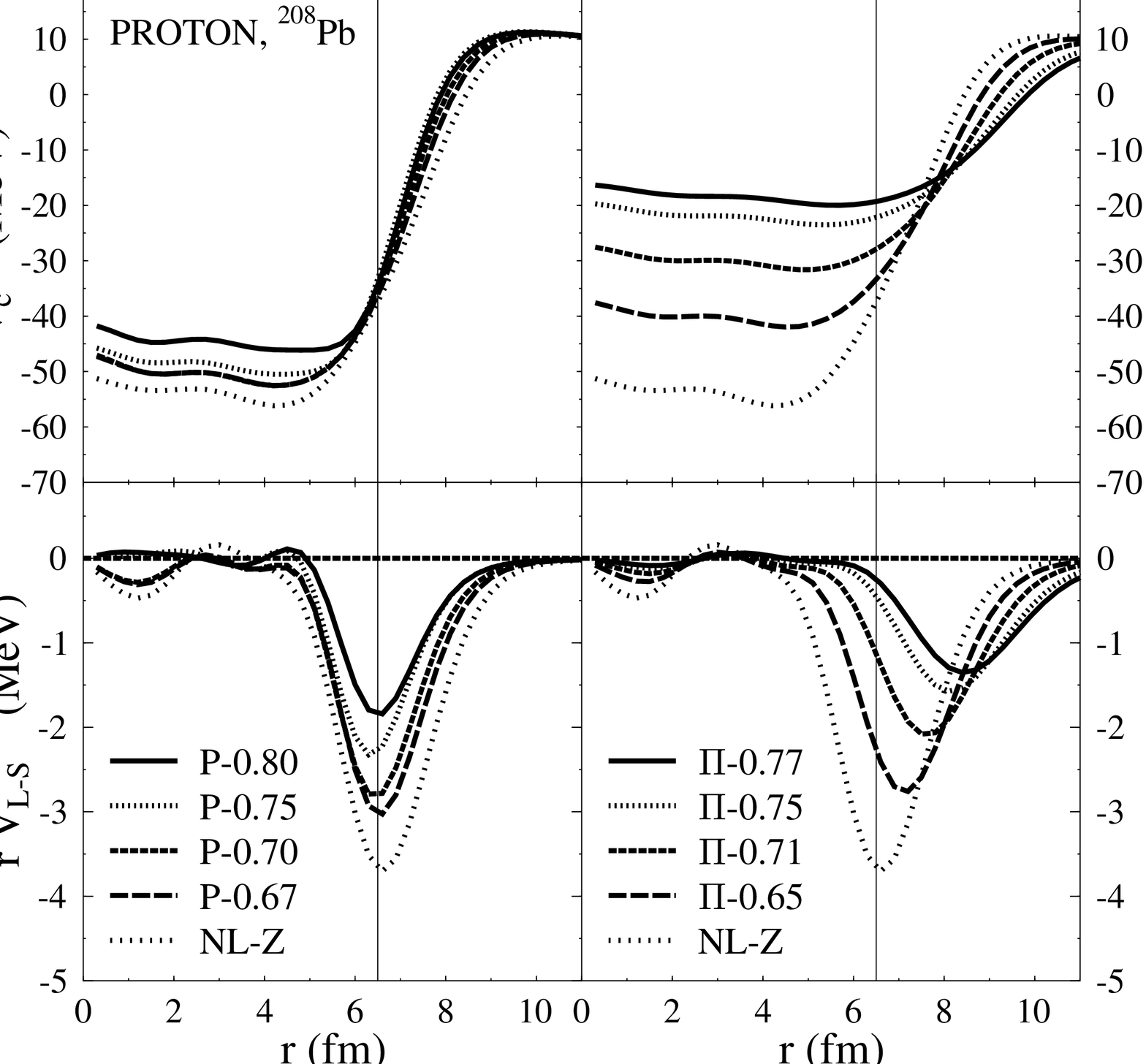,height=7.2 cm}}
\caption{Same as in Fig.~\ref{poteff}, but the  parameter sets are adjusted  with respect to certain  $M^*$. P-067, P-0.70, P-0.75, and P-0.80  are fitted parameter sets. The fitting procedure is the same as in the case of NL-Z. $\Pi$-0.65, $\Pi$-0.70, $\Pi$-0.75 and $\Pi$-0.77 are unfitted parameter sets (obtained only by adjusting the value of coupling constant $g_s$).\label{poteffm*}}
\end{figure}

\begin{figure*}
\centering
 \mbox{\epsfig{file=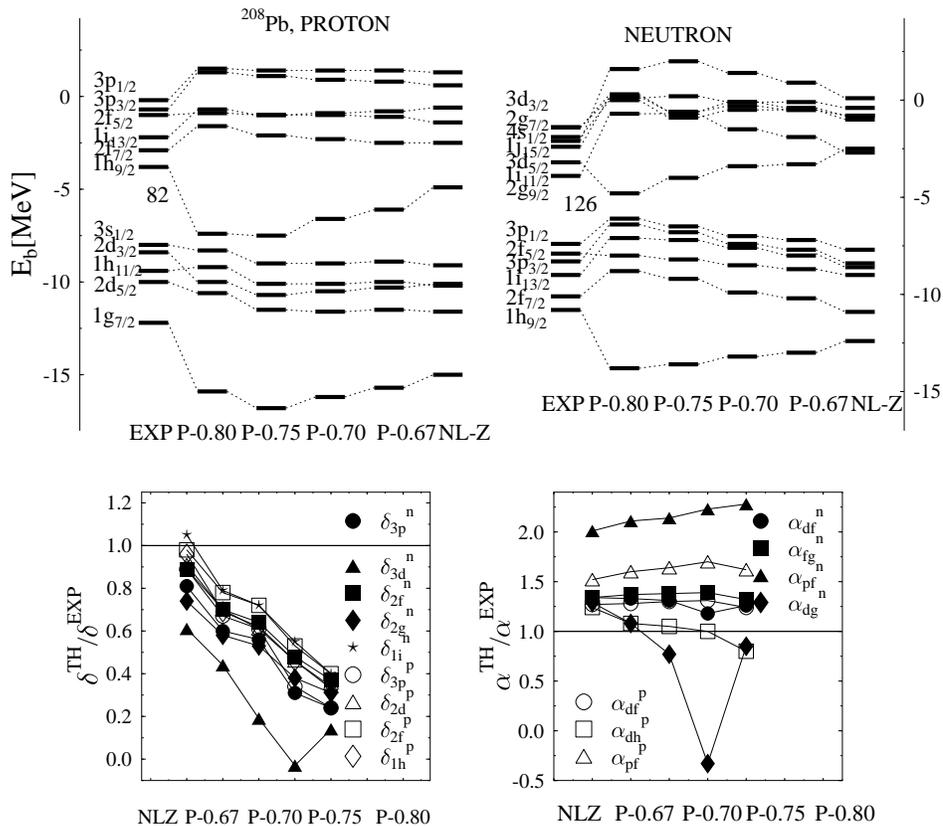,height=11.0cm}}
\caption{Same as in Fig.~\ref{Pb208sps}, but using the P-067, P-0.70, P-0.75, P-0.80 parameter sets.\label{Pb208spsm*}}
\end{figure*}

For the unfitted parameter sets (see the top right panel of
Fig.~\ref{poteffm*}), the depth of $V_{\rm c}^{\rm eff}$ and its trend around Fermi
surface drastically change.  On the other hand, for the fitted parameter sets
(see top left panel of Fig.~\ref{poteffm*}), the depth and the trend around
Fermi surface do not significantly change when $M^*/m$ is varied.  It means that
the nuclear observable requires a cancellation between scalar and
vector potentials. The scalar and vector potentials tend to weaken if
$M^*/m$ becomes larger than 0.6, but the $V_{\LS}^{\rm eff}$ is sensitive
to the variation of $M^*/m$ (see the lower panels of
Fig.~\ref{poteffm*}).  $V_{\LS}^{\rm eff}$ decreases when  $M^*$/m
increases.  The weakening of $V_{\LS}^{\rm eff}$ is due to the smallness
of the spin-orbit splitting (see the upper-left panel of Fig.~\ref{Pb208spsm*}).  Fitting
the parameters to the nuclear observable does not help in this case.  Therefore,
only appropriate values of scalar and vector potentials can yield a  correct
$V_{\rm c}^{\rm eff}$ and  $V_{\LS}^{\rm eff}$ simultaneously. These potentials correspond to an $M^*/m$ of about 0.6.

\begin{figure}[b]
\centering
\mbox{\epsfig{file=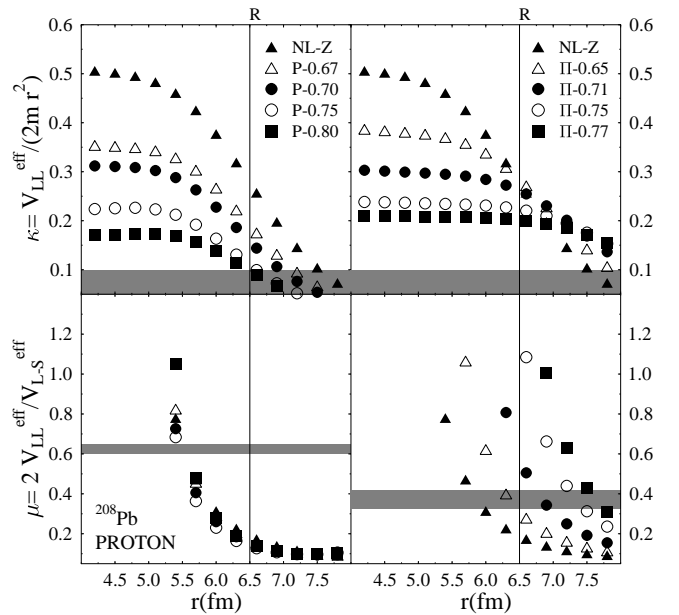,height=7.2 cm}}
\caption{Same as in Fig.~\ref{kamu}, but using the P-067, P-0.70, P-0.75, P-0.80, $\Pi$-0.65, $\Pi$-0.70, $\Pi$-0.75 and  $\Pi$-0.77 parameter sets. \label{kamum*}}
\end{figure}

As can be seen in Fig.~\ref{kamum*}, the Nilsson parameter $\kappa$ depends on
$M^*/m$. The value of $\kappa$ at Fermi surface decreases when  $M^*/m$ increases
and the major effects are found for the fitted parameter sets.  For
these parameter sets, if $M^*$ = 0.77 $m$, it coincides with the
prediction from the Nilsson model (shaded area).  On the other hand, $\mu$ from
fitted parameter sets is almost independent from the $M^*/m$ variation, while $\mu$
of the unfitted parameter sets, on the contrary, depends on $M^*/m$.  It means
that for the RMF models, a constant value of $\mu$ is the requirement for
the correct nuclear bulk properties.  The value of $\mu$ of the RMF models
around Fermi surface is smaller than that of the Nilsson model.  In the RMF models, $\kappa$ and $\mu$ depend on each other. As $M^*/m$ is getting larger, $\kappa$ is getting smaller and the spin-orbit potential is getting smaller in order to
keep $\mu$ constant.

The variation of the neutron and proton SPS with respect to the variation of $M^*$/m
is shown in Fig.~\ref{Pb208spsm*}.  Some states are shifted up and some other
are shifted down as $M^*/m$ increases.  This combination does not
improve the SPS relative position.  The reason is mainly due to the spin-orbit
splitting which is getting narrower when $M^*/m$ increases.  In the
relative position between two different spin-orbit partners ($\alpha$) a couple of states are closer to experimental
data but the majority of states deviate more when $M^*/m$ increases.
The value of $\alpha_{dg}$ is negative for P-070 because there is an exchange of ordering in
$3d$ states (the value of the gap is negative).  Except for $\alpha_{dg}$, all
$\alpha$s change quite drastically.  This effect depends on $M^*/m$, but the
pattern of changing is not the same for every level.  It seems that the
relative position ($\alpha$) can be reproduced if $M^*/m$ depends
on the states.  This dependence can be generated only if we take into account
the exchange and/or other correlations (effects) beyond mean-fields. It would be interesting to see whether or not the exchange effect can remedy this problem. Future calculation should address this question.

\section{Conclusion}
The $\nuc{Pb}{208}{}$, $\nuc{Sn}{132}{}$ and $\nuc{Ca}{40}{}$ SPS of RMF models have been revisited and studied.
Qualitatively, all RMF models presented here have a similar trend in
SPS. Quantitatively, however, they still have differences due to the models.  From
the comparison with  new experimental data~\cite{Isakov}, it is shown that the position of
the state is not only poorly reproduced, but some level positions in neutron spectra are
reversed.

The non-relativistic limit of the RMF model has been derived in which the potentials resemble a Wood-Saxon and Nilsson forms. 
The energy-dependent potentials $V_{\rm c}$ and $V_{LS}$ (in Wood-Saxon type) of RMF models can be transformed into energy-independent potentials $V_{\rm c}^{\rm eff}$ and $V_{LS}^{\rm eff}$ (in Nilsson type) but with an additional angular momentum-dependent potential 
$V_{LL}^{\rm eff}  \boldsymbol{L}^2$. 
These potentials are used to analyze
the $\nuc{Pb}{208}{}$ SPS predictions from several RMF models.  We found that, first, the behavior of  $\kappa$ and $\mu$ of RMF models is different from that of the Nilsson model. Second, due to
the inter-dependence of parameters $\kappa$ and $\mu$ in RMF models, the acceptable
parameter sets ($M^*/m \approx 0.6$) at the Fermi surface need a
relatively large $V_{LL}^{\rm eff}$ in order to maintain a correct spin-orbit
splitting. 

Since the effect of tensor terms in the RMF model is too small in 
$V_{LS}$ of heavy nuclei ($\nuc{Pb}{208}{}$), the effect is marginal to give correct level
 spacings and placement ordering.
The suspicion that a relatively small $V_{LL}^{\rm eff}$ (large $M^*/m$) yields a relatively better placement of states is found to be wrong. It is shown in the $\nuc{Pb}{208}{}$ case that when 
$V_{LL}^{\rm eff}$  is decreasing only two placements of the states are getting better, whereas the rest are getting worse. Therefore, it seems that the problem of RMF models in reproducing experimental data on the relative
placement of the states is originated in the independence of $M^*$ from $\emph{l}$ (state).

\section*{ACKNOWLEDGMENT}
A. S. acknowledges J. A. Maruhn and T. Cornelius for some comments in the early stage of this work.

\end{document}